\documentclass[12pt]{iopart}
\begin{document}
\newtheorem{algorithm}{Algorithm}
\title[The Ultimate Solution to the Quantum Battle of the Sexes game]{The Ultimate Solution\\ \qquad to the Quantum Battle of the Sexes game}
\author{Piotr Fr\c{a}ckiewicz}
\address{Institute of Mathematics of the
Polish Academy of Sciences\\ 00-956, Warsaw, Poland}
\newtheorem{lemma}{Lemma}[section]
\newtheorem{theorem}[lemma]{Theorem}
\newtheorem{example}[lemma]{Example}
\ead{P.Frackiewicz@impan.gov.pl}
\begin{abstract}
We present the unique solution to the Quantum Battle of the Sexes
game. We show the best result which can be reached when the game
is played according to Marinatto and Weber's scheme. The result
which we put forward does not surrender the criticism of previous
works on the same topic.
\end{abstract}
  \maketitle
\section{Introduction}
Theory of games concerns description of conflict situations
between two or more individuals, usually called players.

\noindent Besides classical theory of games for about 10 years has
been a new field of investigation - quantum games \cite{Iqbal}. It
represents an extension of traditional theory of games into the
field of quantum mechanics (quantum information). In quantum games
players have an access to strategies which are not encountered in
the `macroscopic world'. This phenomenon implicates new
interesting results which may be attained by players equipped with
quantum strategies [2, 3, 4, 6 - 12].

\subsection{Battle of the Sexes}

The Battle of the Sexes is a static two-player game of nonzero sum
whose matrix representation is as follows:
\begin{center}
$\begin{array}{l} ~~~~~~~~~~~~~~~~ \quad q = 1  \quad  q = 0 \\
\Gamma: \begin{array}{c} p = 1 \\ p = 0 \end{array}
\left[\begin{array}{cc}
(\alpha,\beta) & (\gamma,\gamma) \\
(\gamma,\gamma) & (\beta,\alpha) \\
\end{array}\right] \quad $where$ \quad \alpha
> \beta > \gamma.
\end{array}$
\end{center}
Characteristic for the Battle of the Sexes game are three Nash
equilibria: one is found in mixed strategies and the other two in
pure strategies. The first player prefers equilibrium $(1, 1)$
which yields him the payoff $\alpha$. In turn the second player,
in order to get the payoff $\alpha$, prefers $(0, 0)$. The problem
of opposing expectations of the two players constitutes definite
dilemma. The players, following their preferences, may play
strategy profile $(1,0)$ that gives them the payoff $\gamma$ - the
least payoff in the game.
\subsection{The model of quantum game}
The quantum model of a two-player static game (the game in which
each player chooses their strategy once and the choices of all
players are made simultaneously) is~a~family $(\mathcal{H},
\rho_{in}, U_{A}, U_{B}, \succeq_{A}, \succeq_{B})$
\cite{Eisert2}. In such a model $\mathcal{H}$ is the underlying
Hilbert space of the physical system used to play a game and
$\rho_{in}$ is the initial state of this system. Sets of
strategies of two players are sets $U_{A}$ and $U_{B}$ of unitary
operators by which players can act on $\rho_{in}$. The symbols
$\succeq_{A}$ and $\succeq_{B}$ mean the preference relation for
the first and the second player, respectively, which can be
replaced by the payoff function. The first scheme for playing
quantum $2\times2$ game in which both players have an access to
`quantum' strategies appeared in \cite{Eisert}. In this model
Hilbert space $\mathcal{H}$ is defined as
$\mathrm{C}^2\otimes\mathrm{C}^2$. Players apply unitary operators
acting on $\mathrm{C}^2$ which depend on two parameters. Their
strategies and the initial state $\rho_{in}$ is taken to be a
maximally entangled state of two qubits. Marinatto and Weber
\cite{MW} introduced a new scheme for quantizing $2\times2$ games.
In contrary to the scheme proposed in \cite{Eisert}, they
restricted players actions to applying identity operator $I$ and
Pauli operator $\sigma_{x}$ or any probabilistic mixture of $I$
and $\sigma_{x}$. This limitation of unitary operators can lead to
the situation in which the players are even unable to state
whether they play a game in classical or in quantum form
\cite{PF}. For this reason Marinatto and Weber's model seems to be
the more natural way for quantizing games. In the next section we
give precise description of this scheme.
\section{General Marinatto-Weber scheme}
In the Marinatto-Weber scheme of playing $2 \times 2$ quantum
games a space state of a game is the $2 \otimes 2$ dimensional
complex Hilbert space with a base $(|00\rangle, |01\rangle,
|10\rangle, |11\rangle)$. The initial state of a game is
$|\psi_{in}\rangle = a_{00}|00\rangle + a_{01}|01\rangle +
a_{10}|10\rangle + a_{11}|11\rangle$ and $I$, $C = \sigma_{x}$ are
unitary operators. Players are able to manipulate the initial
state $|\psi_{in}\rangle$ by employing $I$ or $C$ to the first and
the second entry in the ket $\vert\cdot\cdot \rangle$,
respectively. According to the idea of mixed strategies they can
also apply, respectively, $pI + (1-p)C$, $qI + (1-q)C$. If the
language of density matrices is used then $\rho_{in} =
|\psi_{in}\rangle\langle\psi_{in}|$ and the final state of the
game is as follows:
\begin{eqnarray*}
\rho_{fin} = pqI_{1} \otimes I_{2}\rho_{in}I_{1} \otimes I_{2} +
p(1-q)I_{1} \otimes
C_{2}\rho_{in}I_{1} \otimes C_{2}  \\
+(1-p)qC_{1} \otimes I_{2}\rho_{in}C_{1} \otimes I_{2} +
(1-p)(1-q)C_{1} \otimes C_{2}\rho_{in}C_{1} \otimes C_{2}
\end{eqnarray*}

\noindent When the original classical game is defined by a bi-matrix:\\
\begin{center}
$\begin{array}{c} ~~~~~~~~~~~~~\quad q = 1  ~~~~ \quad q = 0 \\
\Lambda: \begin{array}{c} p = 1 \\ p = 0 \end{array}
\left[\begin{array}{cc}
(x_{11}, y_{11}) & (x_{12},y_{12}) \\
(x_{21}, y_{21}) & (x_{22}, y_{22}) \\
\end{array}\right]

\end{array}$ \end{center}

\noindent the payoff operators are:
\begin{eqnarray*}
P_{A} = x_{11}|00\rangle\langle00| + x_{12}|01\rangle\langle01| +
x_{21}|10\rangle\langle10| + x_{22}|11\rangle\langle11|,\\
P_{B} = y_{11}|00\rangle\langle00| + y_{12}|01\rangle\langle01| +
y_{21}|10\rangle\langle10| + y_{22}|11\rangle\langle11|.
\end{eqnarray*}
The payoff functions $\pi_{A}$ and $\pi_{B}$ can then be obtained
as mean values of the above operators:
\begin{eqnarray*}
\pi_{A} = Tr\{P_{A}\rho_{fin}\}, \quad \pi _{B} =
Tr\{P_{B}\rho_{fin}\}
\end{eqnarray*}
After applying the procedure discussed above, the quantum
equivalent of the classical game $\Lambda$ is characterized by a
two-dimensional bi-matrix $\Lambda_{Q}$, the elements of which are
specified as a product of two matrices:
\begin{eqnarray}
\pi(i,j) = (\begin{array}{cccc}|a_{i \oplus_{2} 1, j \oplus_{2}
1}|^2 & |a_{i \oplus_{2} 1, j }|^2 & |a_{i, j \oplus_{2} 1}|^2 &
|a_{ij}|^2
\end{array}) \bigl(X, Y\bigr)
\end{eqnarray}
where: $\pi(i,j) = (\pi_{A}(i,j), \pi _{B}(i,j))$, $i, j \in
\{0,1\}$, $\oplus_{2}$ means addition modulo 2 and  $\bigl(X,
Y\bigr)$~= $\left(\begin{array}{cccc} (x_{11}, y_{11}) &
(x_{12},y_{12}) & (x_{21}, y_{21}) & (x_{22}, y_{22})
\end{array}\right)^{T}$.\\

\noindent In the special case when $|\psi_{in}\rangle =
|00\rangle$ an equality $\Lambda = \Lambda_{Q}$ occurs.
\section{Various attempts of solving the dilemma of the Quantum Battle of the Sexes game}
The history of efforts put into the quantum solution to the
dilemma that unavoidably occurs in the classical Battle of the
Sexes began in \cite{MW}, where the scheme of playing quantum
games alternative to the scheme proposed in \cite{Eisert} was
published. Marinatto and Weber showed that the players who have an
access to quantum strategies may gain the same payoff in every
equilibrium. If the initial state of the game is $|\psi\rangle =
(|00\rangle + |11\rangle)/\sqrt{2}$ then instead of $(\alpha,
\beta)$ or $(\beta, \alpha)$, respectively, for strategy profiles
$(1,1)$ and $(0,0)$, they obtain $\bigl((\alpha + \beta)/2,
(\alpha + \beta)/2)\bigr)$. Equalization of payoffs for players
obtained in both equilibria certainly eliminates differences
between preferences of the players but, as Benjamin
\cite{Benjamin} correctly stated, the dilemma still exists.
Despite of the fact that both players prefer two equilibrium
situations to the same extent, there is still a possibility that
because of lack of communication between both players they may
obtain the worst payoff $\gamma$, which happens when they play combinations of strategies $(1,0)$ or $(0,1)$. \\
Further improvement in solving the dilemma of the Battle of the
Sexes game was presented by Nawaz and Toor in \cite{NT1}. They
improved the results~of~\cite{MW}~considering the quantum game
Battle of the Sexes that begins with the initial state
$|\psi_{in}\rangle~=~(~\sqrt{5}\e^{i\phi_{1}}|00\rangle +
\sqrt{5}\e^{i\phi_{2}}|01\rangle + \e^{i\phi_{3}}|10\rangle +
\sqrt{5}\e^{i\phi_{4}}|11\rangle$)/4 and showing that it is
equivalent to the classical game characterized by the following
payoff bimatrix:
\begin{center} \mbox{$\Gamma_{NT} =
1/16\left[\begin{array}{cc}
(5\alpha + 5\beta + 6\gamma,5\alpha + 5\beta + 6\gamma) & (5\alpha + \beta + 10\gamma,\alpha + 5\beta + 10\gamma) \\
(\alpha + 5\beta + 10\gamma,5\alpha + \beta + 10\gamma) & (5\alpha + 5\beta + 6\gamma,5\alpha + 5\beta + 6\gamma) \\
\end{array}\right]$} \end{center}
Then they argued that every player should choose their first
strategy. It can be easily observed that \cite{NT1} improve
results of \cite{MW}. For any $\alpha$, $\beta$, $\gamma$ where
$\alpha > \beta
> \gamma$, it is better for both players to play `Nawaz and
Toor's game' than `Marinatto and Weber's game': in \cite{MW}, if
players choose their strategies 1 or 0 at random, they gain with
equal probability $(\alpha + \beta)/2$ and $\gamma$, what gives
them  the expected value $(\alpha + \beta + 2\gamma)/4$ - the
result which is always worse than $(5\alpha + 5\beta +
6\gamma)/16$. However, the question arises: is Nawaz and Toor's
result the best result which players can guarantee themselves in
the quantum Battle of the Sexes game?

\section{Harsanyi - Selten algorithm of equilibrium selection}
The algorithm of choosing equilibrum presented below is described
in a renowned book by Nobel Prize winners Harsanyi and Selten
\cite{HS}. Its aim is to select in each $2\times2$ game with two
strong
equilibria only one of them or the equilibrium in mixed strategies.\\
To demonstrate an operation of the algorithm let us consider the
following $2\times2$ game:
\begin{center}
$\begin{array}{c} ~~~~~~~~~~~~~\quad q = 1  ~~~~ \quad q = 0 \\
\Delta = \begin{array}{c} p = 1 \\ p = 0 \end{array}
\left[\begin{array}{cc}
(a_{11}, b_{11}) & (a_{12},b_{12}) \\
(a_{21}, b_{21}) & (a_{22}, b_{22}) \\
\end{array}\right]
\end{array}$ \end{center}
and denote by $u_{1} = a_{11} - a_{21}$, $v_{1} = a_{22} -
a_{12}$, $u_{2} = b_{11} - b_{12}$, $v_{2} = b_{22} - b_{21}$.
Furthermore, let us assume that the pairs of pure strategies
$(1,1)$, $(0,0)$ form strong equilibria (analogical criterion can
be formulated for equilibria placed on the second diagonal). Then
there exists also the third equlibrium $(s_{1}, s_{2})$ in mixed
strategies, where $s_{1} = v_{2}/(u_{2} + v_{2})$, $s_{2} =
v_{1}/(u_{1} +
v_{1})$.\\
\\
\noindent \textbf{Algorithm:}  \textit{From three equilibria the
one which dominates according to payoffs, i.e the one, in which
both players receive the largest payoffs should be chosen. If this
is not a case, then the equilibrium should be chosen according to
the following formula:
\[\ (r_{1},r_{2}) = \left\{\begin{array}{lll} (1,1), & \mbox{if
$u_{1}u_{2} >
v_{1}v_{2}$} \\
(0,0), & \mbox{if $u_{1}u_{2} <
v_{1}v_{2}$} \\
(s_{1}, s_{2}), & \mbox{if $u_{1}u_{2} = v_{1}v_{2}$}.
\end{array}\right.\]
Such strategy pair is called as risk-dominant equilibrium} \cite{HS}.\\

\noindent It is important to notice that the given algorithm is
not contradictory to individual rationality. The algorithm should
not be treated as an oracle which gives players unjustified hints
which are in conflict with common sense.
The criterion entirely reflects rational behavior of the players (see comments in \cite{HS}). \\

\noindent In order to see how this algorithm works, we apply it to the quantum version of the game the Battle of the Sexes studied by Nawaz and Toor in \cite{NT1} and described by the payoff bi-matrix $\Gamma_{NT}$.\\
It can be easily noticed that the game $\Gamma_{NT}$ has three
equlibria but none of them is dominant according to payoffs. Since
$u_{1} = u_{2} = 4(\alpha - \gamma)$ and $v_{1} = v_{2} = 4(\beta
- \gamma)$, we get $u_{1}u_{2} = 16(\alpha - \gamma)^2 > 16(\beta
- \gamma)^2 = v_{1}v_{2}$. Therefore according to the rule given
by the Harsanyi and Selten's algorithm, players should chose the
equlibrium $(1,1)$ - a strategy pair which also Nawaz and Toor
consider as the only rational solution in this game.
\section{Dilemma of the Battle of the Sexes overcome}

In the previous section we presented the algorithm of equilibrium
selection which should be adapted by rational players for
$2\times2$ games with two strong equilibria. The quantum game
begins when players receive initial state and at this stage there
is a need to define precisely its shape. Below-given lemma will
allow one to use the Harsanyi-Selten algorithm and also to
equalize players' preferences. \\ Let an initial state
$|\psi_{in}\rangle = a_{00}|00\rangle + a_{01}|01\rangle +
a_{10}|10\rangle + a_{11}|11\rangle$ of a quantum $2 \times 2$
game played according to the Marinatto-Weber scheme be given. Then
the original classical game $\Gamma$ transforms into the game
$\Gamma'$ such that the following lemma holds:
\begin{lemma} If $|a_{00}|^2 = |a_{11}|^2 = \frac{1}{2}(1 -
(\epsilon_{1}+ \epsilon_{2}))$, $|a_{01}|^2 = \epsilon_{1},
|a_{10}|^2 = \epsilon_{2}$, where $\epsilon_{1} + \epsilon_{2}
\leq 1 - 2\max\{\epsilon_{1}, \epsilon_{2}\}$ for $\epsilon_{1}
\ne \epsilon_{2}$ and $\epsilon < 1/4$ for $\epsilon_{1} =
\epsilon_{2} = \epsilon$ then for any real numbers $\alpha > \beta
> \gamma$
\begin{enumerate}
\item[a)] a game $\Gamma'$ is identical to  $\Gamma$ with respect
to strategy profiles which constitute Nash equilibria in pure
strategies and with respect to the number of equilibria, \item[b)]
payoff functions $\pi_{A}'$, $\pi_{B}'$ of the quantum game
$\Gamma'$ fulfill the condition: $\pi_{A}'(r_{1},r_{2})
=~\pi_{B}'(r_{1},r_{2})$ for all equilibria $(r_{1},r_{2})$ of the
game $\Gamma'$.
\end{enumerate}
\end{lemma}
\textit{Proof:} \quad Insert $|a_{00}|^2 = |a_{11}|^2 =
\frac{1}{2}(1 - (\epsilon_{1}+ \epsilon_{2}))$, $|a_{01}|^2 =
\epsilon_{1}, |a_{10}|^2 = \epsilon_{2}$ to the formula~(1).
Taking into account assumptions of the lemma about the sum
$\epsilon_{1} + \epsilon_{2}$ we obtain: \begin{eqnarray*}
\pi'_{A}(1,1) - \pi'_{A}(0,1) = \pi'_{B}(1,1) - \pi'_{B}(1,0) =
\\ \bigl(\alpha - \gamma\bigr)\bigl(\frac{1}{2}(1-(\epsilon_{1} +
\epsilon_{2})) - \epsilon_{2}\bigr) + \bigl(\beta -
\gamma\bigr)\bigl(\frac{1}{2}(1 - (\epsilon_{1} + \epsilon_{2})) -
\epsilon_{1}\bigr) > 0. ~~~~(2)
\end{eqnarray*}
Similarly:
\begin{eqnarray*}
\pi'_{A}(0,0) - \pi'_{A}(1,0) = \pi'_{B}(0,0) - \pi'_{B}(0,1) =
\\ \bigl(\alpha - \gamma\bigr)\bigl(\frac{1}{2}(1-(\epsilon_{1} +
\epsilon_{2})) - \epsilon_{1}\bigr) + \bigl(\beta -
\gamma\bigr)\bigl(\frac{1}{2}(1 - (\epsilon_{1} + \epsilon_{2})) -
\epsilon_{2}\bigr) > 0. ~~~~(3)
\end{eqnarray*}
We infer from this results that pairs $(1,1)$, $(0,0)$ form Nash
equilibria and none of the strategies is weakly dominated.
Therefore, the game $\Gamma'$ possesses also an equilibrium in mixed strategies.\\
Furthermore, it can be easily observed that  $\pi'_{A}(1,1) =
\pi'_{B}(1,1) = \pi'_{A}(0,0) = \pi'_{B}(0,0)$. Let us mark by
$(s_{1}, s_{2})$ the third equilibrium of the game $\Gamma'$. Due
to $u_{1} = u_{2}$ and $v_{1} = v_{2}$, we obtain equality
$(s_{1}, s_{2}) = (s_{1}, s_{1}) = (s_{2}, s_{2})$. Therefore,
besides the equalities: $\pi'_{A}(1,0) = \pi'_{B}(0,1)$
and~$\pi'_{A}(0,1) =~\pi'_{B}(1,0)$ we get
$\pi'_{A}(s_{1},s_{2}) = \pi'_{B}(s_{1},s_{2})$.\\

\noindent The essential assumptions of the lemma are not
\textit{condito sine qua non} to fulfill the thesis. Taking into
consideration, for example, another initial state:
$|\psi_{in}\rangle = a_{01}|00\rangle + a_{00}|01\rangle +
a_{11}|10\rangle + a_{10}|11\rangle$ one obtaines a game which is
identical to $\Gamma'$ up to relabelling of strategies of one of
the players. Moreover, the assumption $\epsilon_{1} + \epsilon_{2}
\leq 1 - 2\max\{\epsilon_{1}, \epsilon_{2}\}$ can be weakened. The
asumptions define form of the initial state for the quantum Battle
of the Sexes game with any $\alpha > \beta > \gamma$. The
necessary and sufficient condition for inequalities (2) and (3) to
be true require dependence of $\epsilon_{1}$ and $\epsilon_{2}$ on
$\alpha$, $\beta$ and $\gamma$. However for simplifying the
results, we will not go into details of this problem. As we will
notice further the most important for our study are only values of
$(\epsilon_{1}, \epsilon_{2})$ in the neighborhood of
$(0,0)$. \\
One of the characteristic features of both the classical game the
Battle of the Sexes and any of its quantum versions is the lack of
any equilibria which are dominating according to payoffs. However,
the following theorem states that when assumptions of the lemma
are fulfilled, in quantum version of this game an risk-dominant
equilibrium exists.
\begin{theorem}
If the quantum version $\Gamma'$ of the game $\Gamma$ fulfills
assumptions of the lemma, then its risk-dominant equilibrium is
the strategy profile:
\[\ (r_{1},r_{2}) = \left\{\begin{array}{lll} (1,1), & \mbox{when
$\epsilon_{1} >
\epsilon_{2}$} \\
(0,0), & \mbox{when $\epsilon_{1} <
\epsilon_{2}$} \\
(1/2, 1/2), & \mbox{when $\epsilon_{1} = \epsilon_{2}$}.
\end{array} \qquad \qquad ~~~~~~~~~~~~~~~~~~~~~~~~~~~ (4)\right.\]
\end{theorem}
\textit{Proof:} \quad Let us calculate $u_{1}u_{2}$, and
$v_{1}v_{2}$ from the algorithm and estimate the difference
$u_{1}u_{2} - v_{1}v_{2}$:
\begin{eqnarray*}
u_{1}u_{2} = \left[\bigl(\alpha -
\gamma\bigr)\bigl(\frac{1}{2}(1-(\epsilon_{1} + \epsilon_{2})) -
\epsilon_{2}\bigr) + \bigl(\beta - \gamma\bigr)\bigl(\frac{1}{2}(1
- (\epsilon_{1} + \epsilon_{2})) - \epsilon_{1}\bigr) \right]^2,
\end{eqnarray*}
\begin{eqnarray*}
v_{1}v_{2} = \left[\bigl(\alpha -
\gamma\bigr)\bigl(\frac{1}{2}(1-(\epsilon_{1} + \epsilon_{2})) -
\epsilon_{1}\bigr) + \bigl(\beta - \gamma\bigr)\bigl(\frac{1}{2}(1
- (\epsilon_{1} + \epsilon_{2})) - \epsilon_{2}\bigr) \right]^2,
\end{eqnarray*}
consequently:
\begin{eqnarray*}
u_{1}u_{2} - v_{1}v_{2} = (\alpha + \beta - 2\gamma)(\alpha -
\beta)(1 - 2(\epsilon_{1} + \epsilon_{2}))(\epsilon_{1} -
\epsilon_{2}).
\end{eqnarray*}
The first and the second element of the product is surely
positive. Due to the assumption of the lemma $1 - 2(\epsilon_{1} +
\epsilon_{2})$ is also positive. Therefore, the
sign of the difference $u_{1}u_{2} - v_{1}v_{2}$ depends only on the sign of the difference $\epsilon_{1} - \epsilon_{2}$.\\
In the case when  $\epsilon_{1} =\epsilon_{2}$ the game $\Gamma'$
is characterized by the following equalities:
\begin{eqnarray*}
\pi'_{A}(r_{1}, r_{2}) = \pi'_{B}(r_{1}, r_{2}) \quad \mbox{for
all} \quad (r_{1}, r_{2})
\end{eqnarray*}
\begin{eqnarray*}
\pi'(i,j) = \pi'(i\oplus_{2}1, j\oplus_{2}1) \quad \mbox{for all}
\quad i,j \in \{0,1\}
\end{eqnarray*}
which implicate that Nash equilibrium in mixed strategies is
formed by a pair of strategies $(1/2,
1/2)$.\\

\noindent The initial state is known to the players, so according
to the theorem it determines all the development of the game. The
values of payoff function corresponding to (4) are as follows:
\[\ \pi'_{A,B}(r_{1},r_{2}) = \left\{\begin{array}{lll} \frac{1}{2}\left[(\alpha + \beta) - (\alpha + \beta - 2\gamma)(\epsilon_{1} + \epsilon_{2})\right], &
\mbox{when $\epsilon_{1} >
\epsilon_{2}$} \\
\frac{1}{2}\left[(\alpha + \beta) - (\alpha + \beta -
2\gamma)(\epsilon_{1} + \epsilon_{2})\right], & \mbox{when
$\epsilon_{1} <
\epsilon_{2}$} \\
\frac{1}{4}(\alpha + \beta + 2\gamma) - \frac{1}{2}\epsilon\gamma,
& \mbox{when $\epsilon_{1} = \epsilon_{2} = \epsilon$}.
\end{array}\right.\]
Payoff function depends only on the values of  $\epsilon_{1},
\epsilon_{2}$, thus it can be identified with a~function of two
variables $\epsilon_{1}$ and $\epsilon_{2}$:
\[\ \pi'_{A,B}(\epsilon_{1}, \epsilon_{2}) = \left\{\begin{array}{lll} \frac{1}{2}\left[(\alpha + \beta) - (\alpha + \beta - 2\gamma)(\epsilon_{1} + \epsilon_{2})\right], &
\mbox{when $\epsilon_{1} \ne
\epsilon_{2}$} \\
\frac{1}{4}(\alpha + \beta + 2\gamma) - \frac{1}{2}\epsilon\gamma,
& \mbox{when $\epsilon_{1} = \epsilon_{2} = \epsilon$}.
\end{array}\right.\]
This function is composed of two linear functions with variables
$\epsilon_{1}, \epsilon_{2}$. Let us examine its limit:
\[\ \lim\limits_{(\epsilon_{1}, \epsilon_{2}) \to
(0,0)^{+}}\pi'_{A,B}(\epsilon_{1}, \epsilon_{2}) =
\left\{\begin{array}{lll} \frac{1}{2}(\alpha + \beta), &
\mbox{when $\epsilon_{1} \ne
\epsilon_{2}$} \\
\frac{1}{4}(\alpha + \beta + 2\gamma), & \mbox{when $\epsilon_{1}
= \epsilon_{2} = \epsilon$}.
\end{array}\right.\]
It follows, that:
\begin{eqnarray*}
\sup\limits_{\epsilon_{1}, \epsilon_{2}}\pi'_{A,B}(\epsilon_{1},
\epsilon_{2}) = \frac{1}{2}(\alpha + \beta).
\end{eqnarray*}
The maximum value of the function $\pi'_{A,B}(\epsilon_{1},
\epsilon_{2})$ does not exist, but for any small positive value
$\delta$ an arbiter is able to prepare the initial state with
sufficiently small $\epsilon_{1}, \epsilon_{2}$ that are different
from each other in such way that payoffs of players differ from
$\frac{1}{2}(\alpha + \beta)$ less than $\delta$. This means that
in the quantum Battle of the Sexes game both players may obtain
equal payoffs arbitrary close to $\frac{1}{2}(\alpha+\beta)$.
\begin{example} If $(\alpha, \beta, \gamma) = (5,3,1)$, then
according to the result of Nawaz and Toor each player gets payoff
$2,875$ while our formula yields for an initial state of the game
characterized by $|a_{01}|^2 = \epsilon_{1} = 0,01$, $|a_{10}|^2 =
\epsilon_{2} = 0,02$ and $|a_{00}|^2 = |a_{11}|^2 = 0,485$ payoffs
$[(5 + 3) - 0,03(5+3 - 2)]/2 = 3,91$.
\end{example}
\section{Conclusion}
We obtained a new result in the quantum Battle of the Sexes game
played according to Marinatto-Weber scheme. In contrast to
\cite{MW} we considered the initial state of the game to be most
general state of two qubits. We put conditions for amplitudes of
the initial state so that quantum form of the Battle of the Sexes
game has identical strategic positions of players as the initial
game. Differently from \cite{NT1}, we did not select a particular
initial state, but we examined the dependence of players' payoffs
on amplitudes of base states that form the initial state of the
game. Our research showed that the initial state
$|\psi_{in}\rangle = a_{00}|00\rangle + a_{01}|01\rangle +
a_{10}|10\rangle + a_{11}|11\rangle$ from \cite{NT1} characterized
by $|a_{00}|^2 = |a_{11}|^2 = |a_{01}|^2 = 5/16$, $|a_{10}|^2 =
1/16$ is one of many initial states, which can be prepare without
lost characteristic feature of the classical Battle of the Sexes
game. Moreover, we discovered infinitely more initial states for
which players can achive higher payoffs than by means of Nawaz and
Toor's initial state. This allowed to determine the supremum of
the payoffs values. This quantum version assure that its
participants can get payoffs arbitrary close to the maximal payoff
possible in the game: $\frac{1}{2}(\alpha + \beta)$, which is the
highest value that can be obtained in the `classical' game if and
only if players are allowed to communicate.
\section{Acknowledgments}
The author is very grateful to his supervisor Prof. J. Pykacz from
the Institute of Mathematics, University of Gda\'nsk, Poland for
very useful discussions and great help in putting this paper into
its final form.


\section*{References}
\begin{thebibliography}{10}
\bibitem{Benjamin} S.C. Benjamin `Comment on: A quantum approach to static games of complete information', Phys. Lett. A 277,
180-182 (2000)
\bibitem{Eisert} J. Eisert, M. Wilkens, M. Lewenstein `Quantum Games and Quantum Strategies', Phys. Rev. Lett. 83,
3077-3080 (1999).
\bibitem{Eisert2} J. Eisert, M. Wilkens `Quantum strategies', J.
Mod. Opt. 47, 2543 (2000).
\bibitem{Iqbal} A. Iqbal `Studies in the Theory of Quantum Games', quant-ph/0503176
\bibitem{HS} J. Harsanyi, R. Selten `A General Theory of
Equilibrium Selection in Games', MIT Press, (1988)
\bibitem{MW} L. Marinatto, T. Weber `A Quantum approach to Static games of Complete Information', Phys. Lett.
A 272, 291-303 (2000);
\bibitem{Meyer} D.A. Meyer, `Quantum Strategies' Phys. Rev. Lett. 82,
1052-1055 (1999)
\bibitem{NT1} A. Nawaz, A.H. Toor `Dilemma and quantum battle of
sexes', J. Phys. A: Math. Gen. 37 ~ 4437-4443 (2004)
\bibitem{NT2} A. Nawaz, A.H. Toor `Generalized quantization scheme for two-person non-zero sum
games' J. Phys. A: Math. Gen. 37 11457-11463 (2004)
\bibitem{NT3} A. Nawaz, A.H. Toor `Quantum games with correlated noise', J. Phys. A: Math. Gen. 39 ~ 9321-9328
(2006)
\bibitem{PF} J. Pykacz, P. Fr\c{a}ckiewicz `Arbiter as the Third
Man in classical and quantum games', quant-ph/0707059
\bibitem{Toyota} N. Toyota `Quantization of the stag hunt game and the Nash
equilibrilum', quant-ph/0307029


\end{thebibliography}
\end{document}